\documentclass[prl,reprint,linenumbers,showpacs,amsmath,amssymb]{revtex4-1}
\usepackage{bm}
\usepackage{graphicx}

\newcommand*{\Teff}{T_{\text{eff}}}
\newcommand*{\Tp}{T_{\text{p}}}
\newcommand*{\Tc}{T_{\text{c}}}
\newcommand*{\Tk}{T_{\text{K}}}
\newcommand*{\Te}{T_{\text{E}}}
\newcommand*{\kb}{k_{\text{B}}}
\newcommand*{\Ttft}{T_{\text{TFT}}}
\newcommand*{\eps}{\varepsilon}

\begin{document}

\title{Effective temperatures of a heated Brownian particle}

\date{\today}

\author{Laurent Joly}\email{laurent.joly@univ-lyon1.fr}
\affiliation{LPMCN, Universit\'e de Lyon; UMR 5586 Universit\'e Lyon 1 et CNRS, F-69622 Villeurbanne, France}

\author{Samy Merabia}\email{samy.merabia@univ-lyon1.fr}
\affiliation{LPMCN, Universit\'e de Lyon; UMR 5586 Universit\'e Lyon 1 et CNRS, F-69622 Villeurbanne, France}

\author{Jean-Louis Barrat}\email{jean-louis.barrat@univ-lyon1.fr}
\affiliation{LPMCN, Universit\'e de Lyon; UMR 5586 Universit\'e Lyon 1 et CNRS, F-69622 Villeurbanne, France}

\begin{abstract}
We investigate various possible definitions of an effective temperature for a particularly simple nonequilibrium 
stationary system, namely a heated Brownian particle suspended in a fluid. The effective temperature based on the fluctuation dissipation
ratio depends on the time scale under consideration, so that a simple Langevin description of the heated particle is impossible. The
short and long time limits of this effective temperature are shown to be consistent with the temperatures estimated from the kinetic energy
and Einstein relation, respectively. The fluctuation theorem provides still another definition of the temperature, which is shown
to coincide with the short time value of the fluctuation dissipation ratio.
\end{abstract}


\maketitle

In the recent years, so called ``active colloids'', \textit{i.e.} colloidal particles that exchange with their surroundings
in a non Brownian manner, have attracted considerable attention from the statistical physics community
\cite{Schweizer2003}. These systems are 
of interest as possible models for simple living organisms, and the description of the corresponding nonequilibrium states
 using the tools of  standard statistical physics
raises a number of fundamental questions \cite{Loi2008,Palacci2010}. The most widely studied active colloids are those that exchange momentum 
with the supporting solvent in a non stochastic way, resulting into self propulsion. A less studied possibility
is that the colloid acts as a local heat source and is constantly surrounded by a temperature gradient. 
Experimentally \cite{Radunz2009}, such a situation is achieved when colloids are selectively heated by an external source of radiation which is not absorbed by the solvent.  If the heat 
is removed far away from the particle, or, more practically, if the particle concentration is small enough that the suspending fluid can be considered as a thermostat, a simple nonequilibrium steady state is achieved. 
Each colloidal particle is surrounded by a spherically symmetric halo of hot fluid, and diffuses in an \textit{a priori} Brownian manner.
The diffusion constant of such heated Brownian particles was experimentally shown to be increased compared to the one observed at equilibrium \cite{Radunz2009}, and a semi quantitative analysis of this enhancement was presented in reference \cite{Rings2010}, based on an analysis of the temperature dependence of the viscosity.

In this report, we use simulation to investigate in detail the statistical physics of the simple non 
equilibrium steady state (NESS) formed by a heated particle suspended in a fluid.  The most natural way of describing such a system, in which the particles diffuse isotropically 
in the surrounding fluid, is to make use of a 
Langevin type equation for the center of mass velocity $\bm{U}$, involving in general a memory kernel $\zeta(t)$ and a random force $\bm{R}(t)$:
\begin{equation}
M \frac{\mathrm{d}\bm{U}}{\mathrm{d}t} = - \int_{-\infty}^t  \zeta(t-s) \bm{U}(s) \,\mathrm{d}s  + \bm{F}_{\text{ext}} + \bm{R}(t) .
\label{langevin}
\end{equation}
In a system at thermal equilibrium at temperature $T$, the correlations in the random force and the friction kernel
 are related by the standard fluctuation dissipation theorem, 
$\langle R_\alpha(t) R_\beta(t')\rangle = \delta_{\alpha \beta} \zeta(\vert t-t'\vert ) \kb T$ \cite{Kubo1988}.
Obviously such a description is 
not expected to hold for a heated particle, as the system is now out of equilibrium. A generalization of Eq. \ref{langevin},
involving a corrected fluctuation dissipation relation with an effective temperature $\Teff$ replacing the equilibrium one, would however
 appear as a natural hypothesis. In fact, such an approach was shown to hold for 
sheared systems kept at a constant temperature by a uniform thermostat \cite{McPhie2001}, or in the frame of the particle for a particle driven at constant average speed \cite{Speck2006}. The interpretation of recent experiments \cite{Palacci2010} also makes implicitly use of such a description in describing the sedimentation equilibrium of active particles, or in analyzing the diffusion constant for  hot Brownian motion \cite{Rings2010}.

The use of a Langevin equation with an effective temperature has several direct consequences. The kinetic energy associated with the center of mass, $\langle \frac{1}{2} M \bm{U}^2 \rangle$, is necessarily equal to the effective temperature $\frac{3}{2} \kb \Teff$. 
The diffusion coefficient $D$ and the mobility under the influence of an external force $\mu = U_x/F_x$ are related by an Einstein relation, $D/\mu = \kb \Teff$ \cite{Einstein1905}. More generally, this relation can be seen as the steady state version of the proportionality between the time dependent response function to an external force, $\chi(t) = \delta U_x(t)/\delta F_x$, and the velocity autocorrelation in the nonequilibrium steady state:
\begin{equation}
\chi(t) = \frac{1}{\kb \Teff} \langle U_x(0) U_x(t) \rangle .
\end{equation}
This relation was explored numerically for self propelled particles in reference \cite{Loi2008}, and shown to be consistent with the observed 
Einstein like relation.
Independently of the use of a specific Langevin model, this relation defines an effective temperature trough a so called ``fluctuation dissipation ratio''. The applicability of an effective temperature description is determined by the dependence of this fluctuation dissipation ratio on time. We show in the following that the time scale at which the fluctuation dissipation ratio of a heated particle is determined indeed matters, so that a single temperature description, even in such a seemingly simple system, is problematic. 

Finally, the use of a Langevin description with an effective temperature entails the validity of several ``fluctuation relations'' \cite{Zon2003}, which have been the object of numerous recent experimental and numerical  tests, both in equilibrium and nonequilibrium systems. The study of the fluctuation relation for the heated particle constitutes the last part of this report.

\begin{figure}
\includegraphics[width=0.9\linewidth]{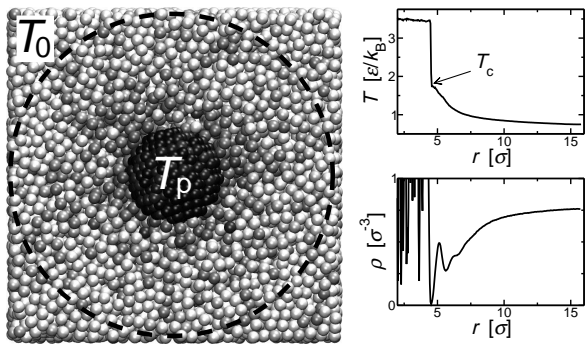}
\caption{\textit{Left}-- Snapshot of the simulated system for $\Tp = 3.5 \eps / \kb$ ($T_0 = 0.75 \eps / \kb$); Gray levels indicate the kinetic energy of atoms. \textit{Right}-- Steady radial temperature and density profiles for this system.}
\label{fig1}
\end{figure}

\begin{figure}
\includegraphics[width=1.0\linewidth]{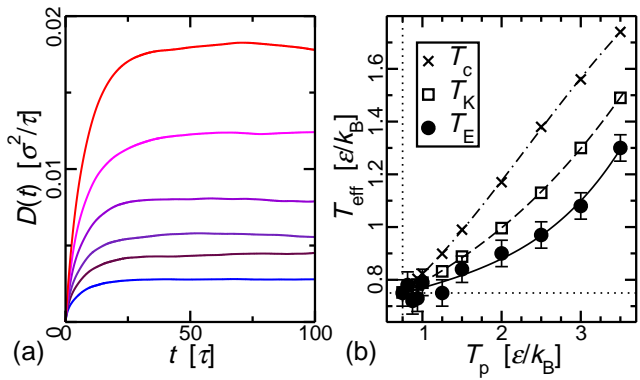}
\caption{(a) Integrated velocity autocorrelation functions of the particle (from bottom to top: $\kb \Tp / \eps = 0.75, 1.5, 2, 2.5, 3, 3.5$). (b) Einstein temperature $\Te$ and kinetic temperature $\Tk$ as a function of the particle temperature $\Tp$; the contact temperature $\Tc$ is also plotted for comparison. Lines are guides for the eye. When not indicated, uncertainties are below the symbol size.}
\label{fig2}
\end{figure}

Our work is based on a direct molecular simulation (MD) approach of a crystalline nanoparticle 
diffusing in a liquid. The simulation were carried out using the LAMMPS package \cite{LAMMPS}.
Details of the model can be found in previous works \cite{Merabia2009,Merabia2009a}, where we used this system to investigate heat transfer from nanoparticles.
The particle was made of 555 atoms with a fcc structure, tied together using FENE bonds. 
The liquid was made of $\sim 23000$ atoms (Fig. \ref{fig1}).
%
%
All liquid and solid atoms interacted \textit{via} the same Lennard-Jones (LJ) potential $v = 4 \eps [ (\sigma/r)^{12}-(\sigma/r)^6]$, at the exclusion of solid atoms directly bonded to each others. In the following, all results will be given in LJ units, namely $\sigma$, $\eps/\kb$ and $\tau = \sqrt{m\sigma^2/\eps}$ for length, temperature and time, respectively.
The atoms in the solid particle  were held at constant temperature $\Tp$ using a Nos\'e-Hoover thermostat, after subtracting the velocity of the center of mass.
In order to mimic the bulk liquid -- far from the particle -- acting as a thermal bath, a rescaling thermostat was applied only to liquid atoms lying beyond $15 \sigma$ from the center of the particle (this condition being evaluated each time the thermostat was applied), to keep them at constant temperature $T_0 = 0.75 \eps/\kb$.
This amounts  to an assumption that the temperature profile around the particle follows the latter instantaneously. This is a reasonable assumption, as  heat diffusion is much faster than mass diffusion in our system: $D_{\text{heat}} \sim 1 \sigma^2/\tau$ \cite{Merabia2009a}, while $D_{\text{mass}} \in [0.002;0.02] \sigma^2/\tau$ (Fig. \ref{fig2}).
Finally the whole system was kept at fixed pressure $p = 0.0015 \eps / \sigma^3$ using a Nos\'e-Hoover barostat.
Simulations were run over typically $10^7$ timesteps in order to accumulate enough statistics.

In  previous work, we have shown that nanoparticles are able to sustain extremely high heat fluxes, \textit{via} two mechanisms: Firstly, interfacial thermal resistance at the nanoscale generates significant temperature jumps at the interface, \textit{i.e.} the contact temperature $\Tc$ of the liquid at the nanoparticle surface is much lower than the particle temperature $\Tp$ (Fig. \ref{fig1}). Secondly, the large curvature-induced Laplace pressure prevents the formation of a vapor layer at the interface; At the highest temperatures, only a stable depleted region is observed (Fig. \ref{fig1}).

Two approaches were used to measure the effective temperature of the particle. We started by measuring the kinetic temperature $\Tk$, related to the center of mass velocity of the  particle.
Due to the finite ratio between solid and liquid masses, care has to be taken to  measure the relative velocity between the solid nanoparticle and the liquid $U_i = U_{\text{s}i} - U_{\text{l}i}$ ($i = x,y,z$), with $U_{\text{s}i}$ and $U_{\text{l}i}$ the velocities of the solid and liquid centers of mass along the $i$ direction.
$\Tk$ was then given by $\frac{1}{2} \kb \Tk = \frac{1}{2} m_{\text{eff}} \langle U_i^2 \rangle$, where $m_{\text{eff}} = m_{\text{s}} m_{\text{l}} / (m_{\text{s}} + m_{\text{l}}$)  [$m_{\text{s}}$ and $m_{\text{l}}$ being the total mass of the  solid and liquid components].
We checked that this procedure behaved correctly for all mass ratios, even when the mass of solid atoms is increased artificially up to the point where $m_{\text{s}} = m_{\text{l}}$. All the velocity measurements presented in the following were done consistently using this procedure.
$\Tk$ was evaluated along the 3 degrees of freedom of the particle in order to estimate the uncertainties, which were  below 1\%.

We also measured the ``Einstein'' temperature $\Te$, defined as the ratio between the diffusion coefficient $D$ and the mobility $\mu$ of the particle \cite{Einstein1905}. 
The diffusion coefficient was computed as the plateau value of the integral of the velocity autocorrelation function (VACF)  $C_{UU}(t) = \langle U_i(t)\,U_i(0) \rangle$ 
of the nanoparticle: $D = \lim_{t \rightarrow \infty} \mathcal{D}(t)$, with $\mathcal{D}(t) = \int_0^t C_{UU}(s) \mathrm{d}s$ (Fig. \ref{fig2}a). The plateau is reached after a correlation time typically around $t_{\text{c}} \sim 30 \tau$.
The mobility $\mu$ was computed by applying an external force $F = 10 \varepsilon / \sigma$ to the particle, and measuring its steady velocity $U$ in the direction of the force: $\mu = U/F$ (linear response in the applied force was carefully checked).

In Fig. \ref{fig2}.b, we have plotted both measures of the particle's effective temperature as a function of $\Tp$.
%
%
One can note that all temperature estimates collapse to $T_0$ at equilibrium. A striking feature of Fig. \ref{fig2}.b is that the two approaches to measure the effective temperature of the particle provide different results. While this is expected for active colloids with a ballistic motion at short times \cite{Palacci2010}, it is quite surprising in the case of a simple Brownian particle, and cannot be understood in the framework of a Langevin description.
As discussed before \cite{Rings2010}, one can finally note that neither $\Tk$ nor $\Te$ identify with the contact temperature $\Tc$, as could be naively expected \cite{Mazo1974} (Fig. \ref{fig2}.b).

To understand the existence of two temperatures in the system, we have probed the fluctuation dissipation theorem (FDT) 
for the Brownian system under study. Generally speaking, considering a physical observable $A$, the response of a system driven out of equilibrium at time $t=0$ 
by the action of a small external field $\mathcal {F}(t)$ is characterized by the susceptibility
$\chi_{AC}(t) = \frac{\langle  \delta A(t) \rangle}{\delta \mathcal {F} (0)}$ where in the subscript of the susceptibility, 
$C$ refers to the variable conjugated to the 
field $\mathcal F$: $C=\frac{\delta \mathcal H} {\delta \mathcal F}$, $\mathcal H$ being the Hamiltonian of the perturbed system.
The FDT states that the susceptibility $\chi_{AC}(t)$ is related to the 
equilibrium correlation function $C_{A C}(t) = \langle A(t) C(0) \rangle$ through: $\int_{0}^{t}\chi_{AC}(s) \mathrm{d}s = \frac{1}{\kb T} C_{AC}(t)$ where $T$ is the thermal bath temperature, and the correlation function is estimated at equilibrium. A sensitive way of probing the deviation from  this relation in nonequilibrium systems, 
which has been extensively used for example in glassy systems \cite{Cugliandolo1997,Berthier2002} consists in determining separately the
integrated  susceptibility  function and the correlation function, and in plotting them in a parametric plot with the time as parameter.
The slope of the curve is then interpreted as the inverse of an  effective temperature, which may depend on the time scale \cite{Cugliandolo1997}.

For the system under study, we obtain the integrated response to an external force $F$  by applying the force in a stationary configuration at $t=0$,
and 
following the evolution of the particle center of mass velocity $U(t)$. The parametric plot involves then 
the average velocity divided by the applied force, $\mu(t) = \langle U(t) \rangle/F = \int_0^t \chi_{UX}(s) \mathrm{d}s$, versus  the integrated velocity auto correlation function $C_{UX}(t) = \int_0^t C_{UU}(s) \mathrm{d}s = \mathcal{D}(t)$. 
To obtain the response function from the ensemble averaged particle velocity $\langle U(t)\rangle$, we have run simulations starting from $1000$ independent configurations of the system and 
tracked the position of the Brownian particle before a steady state is attained (corresponding to times smaller than $t_{\text{c}}$). 
This enabled us to obtain good statistics for the ensemble averaged velocity of the particle, in particular during the early stage of the transient $t \ll t_{\text{c}}$.

%
\begin{figure}
\includegraphics[width=0.8\linewidth]{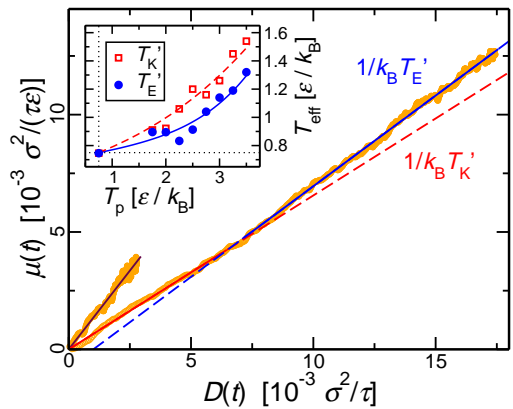}
\caption{Integrated response function as a function of the integrated VACF of the nanoparticle, for $\kb \Tp / \eps = 0.75$ (equilibrium) and $3.5$. \textit{Inset--} Temperatures extracted from the fit of the main graph's curves at small and large times, as a function of the particle
temperature. note that the lines are not merely guides to the eye, but  correspond to the data determined independently and already
reported in  Fig. \ref{fig2}  for the kinetic and Einstein temperature.}
\label{fig3}
\end{figure}
%
Figure \ref{fig3} shows the resulting response/correlation parametric plot, for the different temperatures considered. When $\Tp = T_0$, the nanoparticle is at equilibrium before the external force is applied, and the 
fluctuation dissipation theorem is obeyed. For higher values of the particle temperature $\Tp$, the velocity $\langle U(t) \rangle$ depends non linearly on the integrated VACF
and the fluctuation dissipation ratio is time dependent. This is particularly visible for the highest temperature 
considered in Fig. \ref{fig3} $\Tp = 3.5 \eps/\kb$, where the two slopes $\frac{\mathrm{d} \mu}{\mathrm{d} \mathcal{D}}$ at small and large $\mathcal{D}$ differ markedly.
From these two slopes, it is possible to define 
two temperatures $\Tk'$ and $\Te'$ characterizing the response of the system respectively at short times and long times. The inset of Fig. \ref{fig3}
compares these two  temperatures to the kinetic and Einstein temperatures defined before. Strikingly the short time effective temperature 
$\Tk'$ is very close to the kinetic temperature of the nanoparticle $\Tk$, while the long time effective temperature $\Te'$ is close to the Einstein temperature $\Te$. Therefore our system, in spite of its simplicity, exhibits a ``two temperatures'' behavior on the two different time scales
that are separated by the typical scale set by the loss of memory in the initial velocity. The short time, fast temperature sets the kinetic energy 
of the particles, while the Einstein temperature which probes the steady state response is determined by the long time behavior of the integrated response.

%
\begin{figure}
\includegraphics[width=1.0\linewidth]{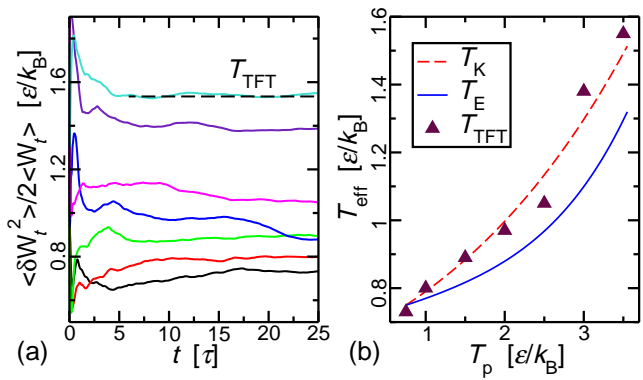}
\caption{(a) Transient fluctuation temperature $T_t = \langle \delta W_t^2 \rangle /2\langle W_t \rangle$ as a function of the time $t$, for different temperatures $\Tp$ of the nanoparticle. From bottom to top: $\kb\Tp/\eps =  0.75, 1, 1.5, 2, 2.5, 3, 3.5$. (b) Transient fluctuation temperature $\Ttft$ 
obtained with the long time limit of $T_t$ as a function of the particle temperature $\Tp$. The lines correspond to the data for the kinetic and Einstein temperature in  Fig. \ref{fig2}.}
\label{fig4}
\end{figure}
%
For a system in contact with a thermal bath and driven out of equilibrium, the bath temperature plays also a key role
in quantifying the  fluctuations of the work from  an external forcing \cite{Zon2003}.
Two situations have to be distinguished depending on the time window analyzed.
If we follow the evolution of a system in the transient regime 
before a steady state is reached, starting from a system at equilibrium, the transient fluctuation theorem (TFT) predicts: 
\begin{equation}
\label{TFT}
P(W_t)/P(-W_t)=\exp(W_t/\kb T) , 
\end{equation}
where $P(W_t)$ is the density probability of the work $W_t$. In this 
equation $W_t$ is the work from the external force $F$, \textit{i.e.} $W_t = \int_{0}^{t} U(s) F \mathrm{d}s$ and $T$ is the  temperature of the thermal bath. On the other hand, in a
a stationary situation, the steady state fluctuation theorem (SSFT) 
predicts $P(W_t)/P(-W_t) \rightarrow \exp(W_t/\kb T)$ when $t \gg t_{\text{c}}$ where $t_{\text{c}}$ denotes a typical equilibrium correlation time.
In the SSFT, the work $W_t$ is estimated along a trajectory of length $t$: $W_t = \int_{t_i}^{t_i+t} U(s) F \mathrm{d}s$, where an average on 
different values of the initial $t_i$ may be performed. We have tested these fluctuation relations for the heated Brownian particles, again applying an external force $F = 10 \eps/\sigma$ 
at $t=0$  and recording the statistics of the work using $1000$ independent configurations. It turned out however that 
the distribution of the work $W_t$ was too noisy to 
determine accurately the ratio $P(W_t)/P(-W_t)$ and critically assess the validity of the fluctuation theorems discussed above.  
To extract an effective temperature measuring the fluctuations of $W_t$, we have used the observation that 
 the statistics of the work $W_t$ is to a good approximation Gaussian. Under these conditions, it is trivial to show
that the distribution of $W_t$ obeys a law similar to Eq. \ref{TFT} with an effective temperature $T_t=\langle \delta W_t^2 \rangle / 2 \langle W_t \rangle$. 
Note that strictly speaking the TFT implies that $T_t=T$ is independent of $t$. In Fig. \ref{fig4}.a we have shown 
the evolution of $T_t$ as a function of the time $t$ for different temperatures $\Tp$ of the nanoparticle. 
For all the temperatures considered, the initially small values of $\langle W_t \rangle$ leads to 
a large uncertainty in the value of $T_t$. For longer times $t > 5 \tau$, the  temperature $T_t$ is approximately independent 
of the time $t$. We will denote $\Ttft(\Tp)$ the value of the effective temperature $T_t$ in this regime. Figure \ref{fig4}.b displays 
the evolution of $\Ttft$ as a function of the temperature of the nanoparticle $\Tp$. It is clear that the resulting $\Ttft$ is very close to 
the kinetic  temperature $\Tk$ characterizing the particle dynamics on short time scales. While we are not aware of a theoretical analysis of this situation, we believe the reason for this proximity lies in the fact that the main contribution to fluctuations in the work function corresponds to the time regime in which the velocity is still correlated to its value at $t=0$, \textit{i.e.} the same time regime in which the fluctuation dissipation ratio corresponds to the ``fast'' temperature.

Our work shows that, even in a conceptually rather simple system, in a nonequilibrium steady state, a description in terms of a Langevin model involving a single temperature is far from trivial. Further generalization and interpretation of the behavior of interacting particles in terms 
of Langevin models and a single noise temperature is expected to suffer similar difficulties, as can already be inferred from the results of  \cite{Loi2008}. 
It would be interesting to explore, if the recent extensions of fluctuation dissipation theorems proposed in refs \cite{Seifert2010,Prost2009} 
can be applied to the present case, \textit{i.e.} to identify observables for which a response-correlation proportionality relation holds. Even so, the 
resulting observables are likely to be different from those that are naturally measured in experiments or simulations. We also note that, with the present observables, experiments using optical tweezers with a strongly absorbing particle could be used to probe the different temperatures investigated here, with the exception of the kinetic one. We expect that such experiments will be able to detect a deviation from equilibrium
of the order of magnitude reported here.

\begin{acknowledgments}
We  acknowledge useful exchanges with L. Bocquet, F. Cichos, K. Kroy and D. Rings, and the support of
ANR project Opthermal.
\end{acknowledgments}


%

\end{document}